\documentclass{mn2e}
\usepackage{graphicx}

\title{{\it Chandra\/} reveals a black-hole X-ray binary within the
ultraluminous supernova remnant MF 16}

\author[T.\,P. Roberts \& E.\,J.\,M. Colbert]
{T.\,P. Roberts$^{1,*}$ \& E.\,J.\,M. Colbert$^2$\\ $^1$ X-ray and
Observational Astronomy Group, Dept. of Physics \& Astronomy,
University of Leicester, University Road, Leicester, LE1 7RH\\ $^2$
Department of Physics and Astronomy, Johns Hopkins University,
Baltimore, MD 21218, USA\\ $^*$E-mail: tro@star.le.ac.uk}
\date{}

\def\ro{{\it ROSAT~\/}}
\def\asca{{\it ASCA~\/}}
\def\ein{{\it Einstein~\/}}
\def\xmm{{\it XMM-Newton~\/}}

\def\hst{{\it HST~\/}}
\def\chan{{\it Chandra~\/}}

\def\ergsec{{\rm ~erg~s^{-1}}}

\def\atpcm{{\rm ~atoms~cm^{-2}}}
\def\ctsec{{\rm ~count~s^{-1}}}

\def\H0{{\rm ~km~s^{-1}~Mpc^{-1}}}

\def\la{\mathrel{\hbox{\rlap{\hbox{\lower4pt\hbox{$\sim$}}}{\raise2pt\hbox{$<$}}}}}
\def\ga{\mathrel{\hbox{\rlap{\hbox{\lower4pt\hbox{$\sim$}}}{\raise2pt\hbox{$>$}}}}}

\def\d25{D$_{25}$}
\def\nh{{$N_{H}$}}
\def\Ha{{H$\alpha$}}
\def\hi{H{\small I}$~$}

\def\.25{0.25 keV\thinspace}

\begin{document}

\maketitle

\begin{abstract}
We present evidence, based on \chan ACIS-S observations of the nearby
spiral galaxy NGC 6946, that the extraodinary X-ray luminosity of the
MF 16 supernova remnant actually arises in a black-hole X-ray binary.
This conclusion is drawn from the point-like nature of the X-ray
source, its X-ray spectrum closely resembling the spectrum of other
ultraluminous X-ray sources thought to be black-hole X-ray binary
systems, and the detection of rapid hard X-ray variability from the
source.  We briefly discuss the nature of the hard X-ray variability,
and the origin of the extreme radio and optical luminosity of MF 16 in
light of this identification.
\end{abstract}


\section{Introduction}

An observational definition of ultraluminous X-ray sources (ULX) is
those X-ray sources detected coincident with nearby galaxies, that are
both located outside the galactic nucleus and display an observed
luminosity in excess of $10^{39} \ergsec$.  There has recently been a
lot of attention paid to this class of object based on the argument
that if they are accretion-powered systems, then in many cases the
observed X-ray luminosity exceeds the Eddington luminosity for a 10
M$_{\odot}$ black hole.  Explanations for this apparent breach of the
Eddington limit include accretion on to a new class of $10^2 - 10^5$
M$_{\odot}$ {\it intermediate-mass\/} black hole (e.g. Colbert \&
Mushotzky 1999; Miller \& Hamilton 2002), super-Eddington radiation
from bloated accretion discs in X-ray binaries (Begelman 2002), and
anisotropic emission from X-ray binaries (e.g. King et al. 2001;
K{\"o}rding, Falcke \& Markoff 2002; King 2002).

However, not all ULX are accreting systems.  A minority are actually
recent supernovae, such as SN 1986J (Bregman \& Pildis 1992), SN 1979C
(Immler, Pietsch \& Aschenbach 1998), SN 1978K (Ryder et al. 1993) and
SN 1995N (Fox et al. 2000) to name but four.  Such objects have been
observed to reach X-ray luminosities of $10^{41} \ergsec$ (e.g. SN
1988Z, Fabian \& Terlevich 1996) and are thought to be the result of
the supernova exploding into a dense circumstellar environment.  The
ULX phase for young supernova remnants can be short, however, with
many (though not all) showing steep X-ray flux decay curves (Immler \&
Lewin 2002).  Reports of much older, evolved supernova remnants with
ULX-like fluxes, for instance the ``hypernova'' remnants of M101 (Wang
1999), have since turned out to be mis-identifications of X-ray
binaries (Snowden et al. 2001).

One supernova remnant that appears to have remained unusually X-ray
luminous whilst becoming quite evolved is the remnant MF 16 in NGC
6946 (nomenclature from Matonick \& Fesen 1997).  Though the X-ray
source was detected in an \ein IPC observation of NGC 6946 (Fabbiano
\& Trinchieri 1987), it was first identified as an extremely X-ray
luminous supernova remnant (L$_{\rm X} \sim 3 \times 10^{39} \ergsec$
at 5.1 Mpc, placing it firmly in the ULX regime) on the basis of \ro
PSPC data (Schlegel 1994).  The X-ray data appeared consistent with a
young supernova remnant, but optical data revealed the remnant to be
evolved, possibly $\sim 3500$ years old (Blair \& Fesen 1994).  The
remnant appears extraordinarily luminous in X-rays, optical line
emission and radio continuum emission (van Dyk et al. 1994) for a
supernova remnant of this age.  Subsequent \hst observations revealed
MF 16 to have a physical size of $\sim 34 \times 20$ pc, and a
distinct morphology characterised by three ``loops'' apparent in
narrow-band \Ha~ and [S{\small II}] images (Blair, Fesen \& Schlegel
2001; hereafter BFS01), though an [O{\small III}] image suggested a
more elliptical structure with a possible young star at the heart of
the nebula.  In conjunction with optical spectral data, BFS01 suggest
that the extraordinary luminosity of this remnant is due to the blast
wave of a young supernova impacting on the dense shell of an older
remnant. \ro and \asca data were used to investigate the X-ray
emission in detail by Schlegel, Blair \& Fesen (2000), who found the
spectral data to fit well to a dual Raymond-Smith thermal plasma
model.  This is consistent with the \hst data if the hot component
comes from the interaction zone and the soft component from within the
supernova bubble.  An alternate view is offered by Dunne, Gruendl \&
Chu (2000), who propose that the remnant originates in the interaction
of a single supernova blast wave with a dense circumstellar nebula
ejected from a massive progenitor star.  They suggest that a hard
component in the \ro PSPC spectral data may orginate in a pulsar wind
nebula.

The first of the two \chan observations reported in this letter has
been presented by Holt et al. (2003).  They conclude that the
featureless X-ray spectrum of MF 16 excludes the interaction of a
blast wave with circumstellar matter, and does not appear to fit to a
pulsar wind nebula model either.  They suggest that the X-ray emission
may originate in an unusual, if not unique, set of circumstances.
Here, we re-analyse this data in tandem with a later public
observation of NGC 6946, and demonstrate that the X-ray source
associated with MF 16 is in fact a black-hole X-ray binary (BHXRB),
explaining the apparent dichotomy of an evolved remnant with a young
supernova-like X-ray flux.

\section{Observations \& data reduction}

\begin{table}
\caption{Details of the \chan observations.}
\begin{tabular}{lccc}\hline
Sequence & Aimpoint & Date	& Exposure \\
number  & (J2000)	&	& (ks) \\\hline
500109	& $20^h34^m49.0^s, +60^{\circ}09'19''$	& 2001-09-07
& 58.3 \\
500406	& $20^h34^m51.2^s, +60^{\circ}09'16''$	 & 2002-11-25
& 30.0 \\\hline
\end{tabular}
\label{obsdetails}
\end{table}

NGC 6946 has been observed by \chan on two separate occasions.  The
details of the observations are given in Table~\ref{obsdetails}.  Both
observations were pointed at its nuclear region, though the target of
the latter observation was the recent supernova SN2002HH, and the
large majority of the area of the galaxy falls on the ACIS-S3 chip.
MF 16 lies about 2.5 arcmin north-east of the observation aimpoint,
fortuitously on the S3 chip in both datasets.

The data was reduced and subsequently analysed using the {\small CIAO
v2.3} and {\small HEAsoft v5.2} packages.  The level 2 event lists
were initially filtered down to an energy range of 0.3 -- 10 keV, and
images were created at the full detector spatial resolution (1 pixel
$\equiv 0.492$ arcsec).  Source spectra and lightcurves were then
extracted using the {\small PSEXTRACT} and {\small LIGHTCURVE} tools
respectively, from within a 10-pixel radius circular aperture set to
encompass the full point spread function of the X-ray source.
Background regions were extracted from a similar-size aperture
displaced by $\sim 10$ arcsec from MF 16.  Neither event list was
time-filtered for background flaring, as the total background measured
from the unfiltered background regions amounted to less than 0.5\% of
the counts accumulated from the source in either observation, allowing
us to exploit the full exposure time.

\section{Evidence that MF 16 contains an X-ray binary system}

The X-ray counterpart of MF 16 (hereafter NGC 6946 X-11, after Roberts
\& Warwick 2000) was detected in both observations as the brightest
X-ray source on the \chan ACIS-S3 chip, at a position
$20^h35^m00.75^s, +60^{\circ}11'30.9''$\footnote{This is an average of
the two positions from the separate observations, though the
difference between the measured positions was a mere $0.4''$.}.  A
total of $8585 \pm 94$ and $3838 \pm 63$ counts were detected from the
source in the 2001 September and 2002 November observations
respectively.  This is currently amongst the largest datasets for an
ULX in the \chan era.  In the following sections we present the
details of its X-ray characteristics.

\subsection{Spatial}

We test whether NGC 6946 X-11 is spatially extended by comparing its
radial profile from the 2001 September observation to a modelled point
spread function (PSF) for a point-like X-ray source at the same
detector position and off-axis angle\footnote{Combining the data from
both observations to produce a higher quality radial profile is not an
option, since the source is at different off-axis angles and detector
positions in the two observations.}.  The model is interpolated from a
library of model PSFs using the {\small MKPSF} tool, at an energy of
1.5 keV.  We account for known residual aspect correction errors
(c.f. section 2.2 of Zezas et al. 2002) by convolving the model PSF
with a 2-D Gaussian function of HWHM $= 0.75$ arcsec.  Radial profiles
for the model and source data were extracted in one-pixel width annuli
around the source centroid, and the model PSF was normalised to the
same central peak as the source.  Approximating the PSF shape by a
Gaussian function showed the source data and the modelled PSF to match
extraordinarily well, with the best fit to each giving a HWHM of $1.31
\pm 0.02$ arcsec for the data, and $1.33 \pm 0.02$ arcsec for the
model (the fitting was performed using a $\chi^2$ minimization
algorithm, and the errors quoted are 90\% errors for one interesting
parameter).  The observed X-ray emission is therefore fully consistent
with originating in a point-like source.  However, by comparison with
Roberts et al. (2002) we note that on-axis Gaussian fits to point-like
sources give a HWHM of $\sim 1$ arcsec for a spatial resolution of 0.5
arcsec.  Scaling this to our observation implies it is sensitive to
structure on scales of $> 0.7$ arcsec, ruling out extended X-ray
emission on the full scale of MF 16 ($1.4 \times 0.8$ arcsec$^2$;
BFS01), though some spatial extension on scales less than this cannot
be excluded.

\subsection{Spectral}

Before undergoing spectral analysis, the auxilliary response files for
both datasets were corrected for the known quantum efficiency
degradation of the ACIS-S3 chip using the {\small CORRARF} tool.  The
extracted source spectra were binned to a minimum of 20 counts per bin
before analysis in {\small XSPEC v11.2}, and only data in bins located
above an energy of 0.5 keV were considered.  The observed 0.3 -- 10
keV count rates of $\sim 0.12 - 0.14 \ctsec$ were accumulated in a
full-frame exposure mode in both observations which, even allowing for
their off-axis positions, implies that the source data will have
suffered from a moderate level ($\sim 10 - 15\%$) of pile-up.  We
account for this in the following analysis using the {\small XSPEC
v11.2} parameterisation of the Davis (2001) CCD pile-up mitigation
algorithm.

\begin{table*}
\caption{Two-component fits to the \chan X-ray spectral data.}
\begin{tabular}{lccccccccc}\hline
Model$^a$	& Epoch$^b$	& $\alpha^c$	& \nh	&
$kT_1$	& Abundance  & $kT_{in}$ & $\Gamma$	& $kT_2$	&
$\chi^2$/dof \\
	& 	&	& ($\times 10^{22} \rm ~cm^{-2}$)	&
(keV)	& (Solar units)	& (keV)	&	& (keV)	& \\\hline
PU*WA*(D+PO)	& A	& $0.16^{+0.06}_{-0.04}$
& $0.37^{+0.04}_{-0.02}$	& -	& -	& $0.16 \pm 0.01$	&
$2.53^{+0.09}_{-0.14}$	& -	& 178.0/155 \\
	& B	& $< 0.19$	& $0.47^{+0.1}_{-0.08}$	& -	& -
& $0.14 \pm 0.01$	& $2.28^{+0.18}_{-0.21}$	& -	&
127.9/112 \\ 
PU*WA*(B+PO)	& A	& $0.16^{+0.06}_{-0.04}$	&
$0.38^{+0.09}_{-0.06}$	& $0.12^{+0.01}_{-0.02}$	& -	& -
& $2.46^{+0.07}_{-0.21}$	& -	& 183.9/155 \\
	& B	& $< 0.31$	&
$0.44^{+0.05}_{-0.11}$	& $0.12 \pm 0.01$	& -	& -
& $2.23 \pm 0.22$	& -	& 128.3/112 \\
PU*WA*(M+PO)	& A	& $0.38^{+0.04}_{-0.08}$	& 0.2$^f$	&
$0.66^{+0.13}_{-0.07}$	& $> 7.4$	& -	& $2.49^{+0.05}_{-0.06}$
& -	& 183.4/155 \\
	& B	& $0.30^{+0.07}_{-0.15}$	& 0.2$^f$	&
$0.77^{+0.08}_{-0.15}$	& $5.7^{+3.0}_{-4.0}$	& -	&
$2.32^{+0.1}_{-0.13}$	& -	& 123.3/112 \\
PU*WA*(M+M)	& A	& $0.42^{+0.04}_{-0.02}$	& 0.2$^f$
& $0.33 \pm 0.02$	& $0.07 \pm 0.01$	& -	& -	&
$2.15 \pm 0.12$	& 200.6/155 \\
	& B	& $0.34^{+0.07}_{-0.02}$	& 0.2$^f$
& $0.64 \pm 0.05$	& $0.14^{+0.01}_{-0.04}$	& -	& -
& $2.43^{+0.78}_{-0.69}$	& 135.4/112 \\
\hline
\end{tabular}
\begin{tabular}{l}
Notes: $^a$ Model components are: PU - pileup correction; WA - cold
absorption; D - multicolour disc blackbody; PO - powerlaw continuum;\\ B
- classical blackbody; M - MEKAL optically thin thermal plasma model.
$^b$ Epoch of observation, with A $\equiv$ 2001 September and B
$\equiv$ 2002\\ November.  $^c$ Grade migration parameter in the
pile-up mitigation model.  $^f$ Parameter value fixed at that shown.
\end{tabular}
\label{specfits}
\end{table*}

\begin{figure*}
\centering
\includegraphics[width=5.2cm,angle=270]{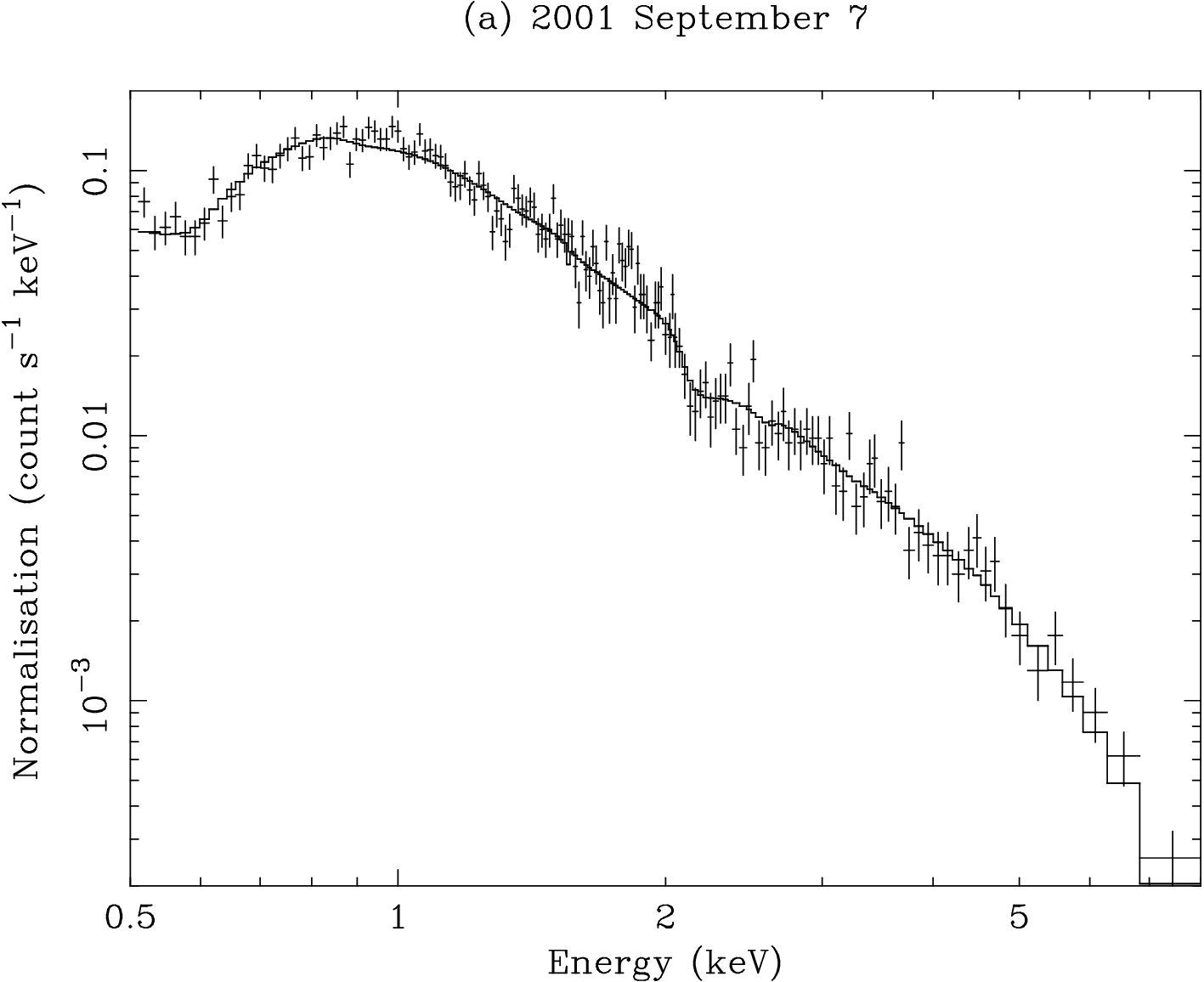}\hspace*{2cm}
\includegraphics[width=5.2cm,angle=270]{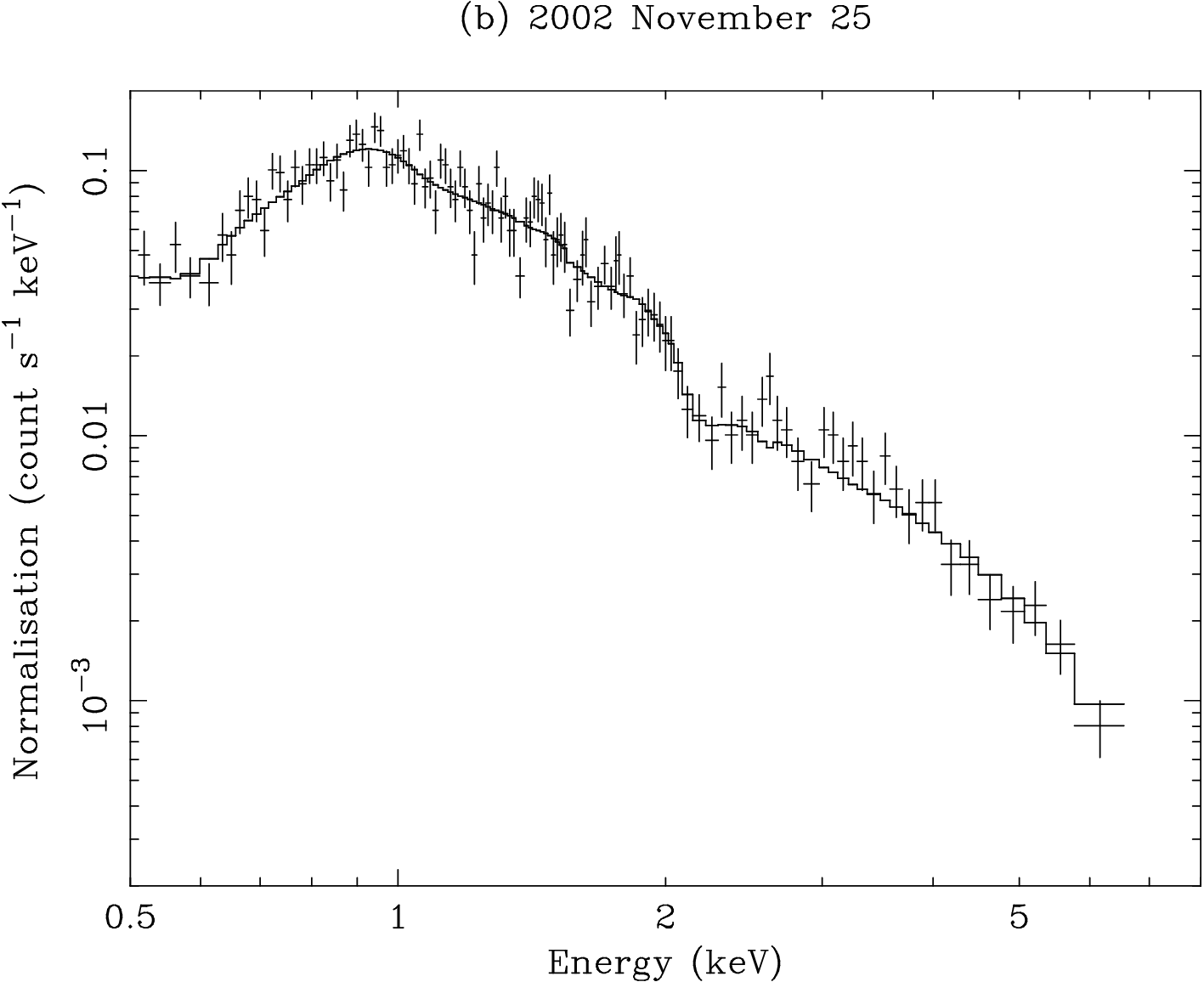}
\caption{The observed X-ray spectra (data points) and best fitting
models (solid lines; c.f. Table~\ref{specfits}) from the \chan
observations of NGC 6946 X-11.  The datasets are plotted on identical
axes for direct comparison.}
\label{specfig}
\end{figure*}

Neither dataset was well-fit by a variety of single-component simple
models, with an absorbed powerlaw continuum the only model to have a
reduced-$\chi^2$ value below 2 for both datasets.  We note that in
both cases a correction for pile-up was justified since it provided an
improvement to the fit at a significance of $> 99.8\%$ for one extra
degree of freedom, though the resulting fits were still poor
($\chi^2$/dof = 239.8/157 and 159.2/114 for the 2001 September and
2002 November observations respectively).

The fits to the data were greatly improved by the addition of a second
spectral component to the absorbed powerlaw continuum model.  We show
the models and best fit parameters in Table~\ref{specfits}.  All
errors are quoted as the 90\% error for one interesting parameter.
The additional component required to improve the fit turned out in all
cases to be a very soft component, which could be modelled by either a
multicolour disc blackbody (hereafter MCDBB; Mitsuda et al. 1984) with
an inner-disc temperature of $kT_{in} \sim 0.15$ keV, a classical
blackbody emission spectrum of temperature $kT \sim 0.12$ keV, or a
MEKAL optically thin thermal plasma model with $kT \sim 0.7$ keV.  The
resulting goodness of fit (as measured by the $\chi^2$/dof) was
remarkably similar for all three models in each epoch, with the model
containing the MCDBB component marginally preferred in the first
epoch, and the model including the MEKAL in the second.  We show the
data and the corresponding best-fitting model for both observations in
Figure~\ref{specfig}.  We note, however, that to constrain a best fit
using the MEKAL plus powerlaw model the line-of-sight column densities
had to be fixed to a minimum value equal to the foreground column
through our own galaxy ($\sim 2 \times 10^{21} \atpcm$; interpolated
from the \hi maps of Stark et al. 1992).  This seems physically
implausible, since the observation of no absorption intrinsic to MF 16
is at odds with the total extinction of $E(B-V) = 0.65$ observed for
its optical emission lines (BFS01).  This extinction converts to a
column of $4.3 \times 10^{21} \atpcm$ (c.f. Zombeck 1990).

\begin{figure*}
\centering
\includegraphics[width=5.4cm,angle=270]{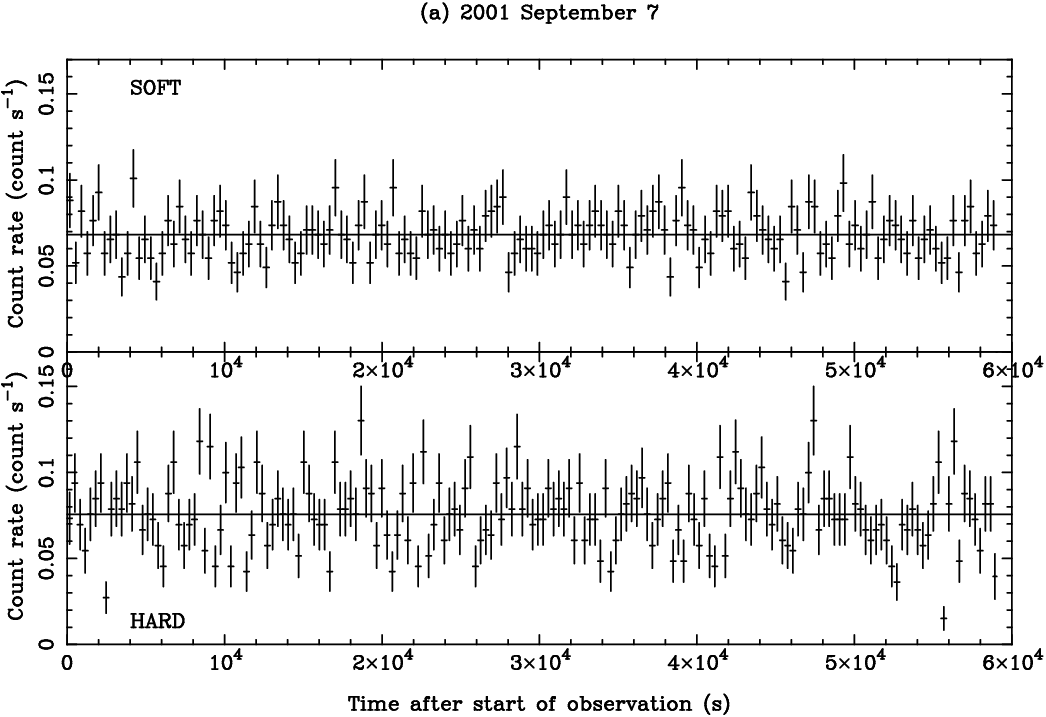}\hspace*{1cm}
\includegraphics[width=5.4cm,angle=270]{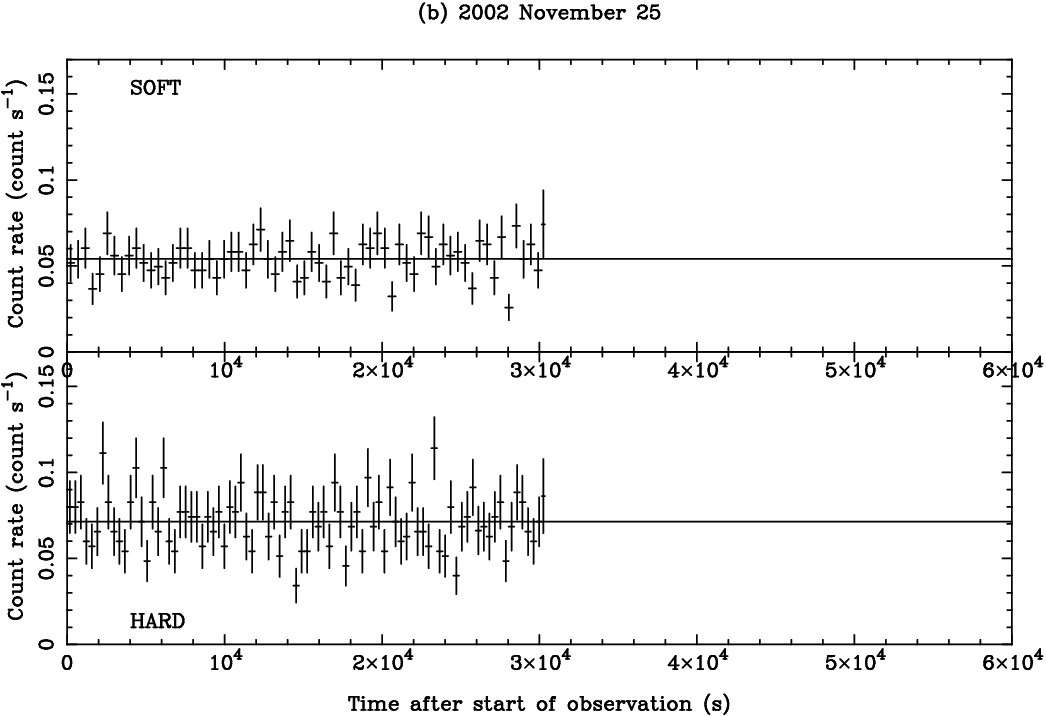}
\caption{\chan lightcurves for NGC 6946, split into soft (0.3 -- 1.1 
keV; top panel) and hard (1.1 -- 10 keV; lower panel) X-ray emission.
Each panel is plotted on identical axes for direct comparison.}
\label{lcs}
\end{figure*}

On the other hand, the MCDBB or classical blackbody plus powerlaw
continuum model fits are consistent with the inferred column to the
central regions of MF 16.  Crucially, the MCDBB plus powerlaw
continuum model has recently been found to describe the \chan or \xmm
X-ray spectra of several ULX that are known or suspected BHXRB
(e.g. Miller et al. 2003).  Furthermore, the derived parameters, and
in particular the inner-disc temperature and normalisation
(constrained at $870^{+660}_{-170}$ and $1600^{+1700}_{-1300}$
km$^{-2}$ (10 kpc)$^{-2}$ in epochs A and B respectively) of the MCDBB
appear extraordinarily similar when our fits are compared to the ULX
studied by Miller et al. (2003).  This suggests that the X-ray
emission processes in NGC 6946 X-11 are those of a BHXRB.  The data
also suggest that the spectrum changes between observations, though as
the error ranges for each of the parameters overlap none of the
changes are very statistically significant.  This tentative suggestion
of spectral variation is also indicative of BHXRB behaviour.  We note
that the observed 0.5 -- 8 keV luminosity of NGC 6946 X-11 from this
model is $\sim 2.5 \times 10^{39} \ergsec$, of which $\sim 20\%$
originates in the MCDBB component.

Finally, we also fit a dual MEKAL model to the data, similar to the
dual Raymond-Smith plasma model used to provide a fit to the combined
\ro PSPC/\asca spectral data (Schlegel, Blair \& Fesen 2000).  We are
again required to constrain the absorption column to the foreground
galactic value.  Our fits provide reasonably similar results to the
low-Z \ro PSPC/\asca fit, albeit with our substantially lower fit
abundances allowing the temperatures of both components to be lower in
the \chan data.  However, this model is clearly the least preferred of
the four two-component spectral models for both \chan datasets,
arguing that the blast wave interaction interpretation of Schlegel,
Blair \& Fesen (2000) is not favoured by the \chan data.

\subsection{Temporal}

Previous analyses of this object have suggested that the X-ray flux
remains constant over a timescale of years (e.g. Holt et al. 2003), as
would be expected for the X-ray emission from an object up to 30 pc in
diameter.  Here, we detect a drop of $\sim 13\%$ in the 0.3 -- 10 keV
count rate over a year, from $0.144 \ctsec$ to $0.125 \ctsec$, clearly
inconsistent with such an extended X-ray source.  However, the \chan
CCDs have a known problem with their quantum efficiency, and hence
effective area to incoming photons, degrading with time.  This could
lead to drops in the detected count rate for separate observations of
constant X-ray fluxes, particularly for sources with soft X-ray
spectra.  We checked this by running simulations in {\small XSPEC},
using the best fit model to the 2001 September data, and the auxiliary
response file (corrected to the degraded area) from the 2002 November
observation.  An overall drop of $\sim 3\%$ in the count rate was
predicted, clearly smaller than the observed deficit, implying that
the source displays at least small amplitude variability on long
timescales.

\begin{table*}
\caption{Tests of the short-term variability of NGC 6946 X-11.}
\begin{tabular}{lcccccccc}\hline
Epoch	& Band$^a$	& Bin size	& \multicolumn{3}{c}{Standard
deviation, $\sigma$}	& \multicolumn{2}{c}{$\chi^2$ statistic}	& K-S
statistic \\ 
	&	& (s)	& Expected$^b$	& Observed$^b$	&
$P_{\sigma}$(var) & $\chi^2$/dof	& $P_{\chi^2}$(var)	&
$P_{\rm K-S}$(var) \\\hline 
2001 September	& Full	& 174	& $2.88 \times 10^{-2}$	& $(3.37 \pm
0.18) \times 10^{-2}$	& $99.7\%$	& 534/340	& $> 99.9\%$
& $\sim 94\%$ \\
	& Soft	& 367	& $1.36 \times 10^{-2}$	& $(1.38 \pm
0.11) \times 10^{-2}$	& -	& 145/160	& -	& - \\
	& Hard	& 330	& $1.51 \times 10^{-2}$	& $(1.97 \pm
0.15) \times 10^{-2}$	& $99.9\%$	& 402/178	& $> 99.9\%$
& $> 99\%$ \\ 
2002 November	& Full	& 199	& $2.51 \times 10^{-2}$	& $(2.65 \pm
0.22) \times 10^{-2}$	& -	& 158/152	& - & - \\
	& Soft	& 464	& $1.08 \times 10^{-2}$	& $(1.03 \pm
0.13) \times 10^{-2}$	& -	& 69/65	& -	& $> 99\%$ \\
	& Hard	& 351	& $1.43 \times 10^{-2}$	& $(1.55 \pm
0.17) \times 10^{-2}$	& -	& 110/86	& $\sim 96\%$
& $\sim 95\%$ \\\hline
\end{tabular}
\begin{tabular}{c}
Notes: $^a$ as defined in the text.  $^b$ in count s$^{-1}$.  We only
show $P$(var) in the instances where it highlights the variability.
\end{tabular}
\label{stats}
\end{table*}

The X-ray variability was further investigated by deriving short-term
lightcurves from both observations.  Both lightcurves were extracted
in the full 0.3 -- 10 keV range and binned to a temporal resolution
equivalent to 25 counts per bin for a constant source count rate, in
order to maximise the temporal resolution whilst maintaining a
reasonable average signal-to-noise ratio of five per bin.  Two simple
tests for variability were then applied to these lightcurves.
Firstly, a classic standard deviation of the count rate per bin from
the mean count rate over the course of the observation was calculated,
and then compared to the expected deviation of $\pm 20\%$ expected
from Gaussian counting noise.  Secondly, a $\chi^2$ test was performed
against the hypothesis that the source count rate remained constant
during the observation.  The results of these tests, including a
probability that the data is variable, $P$(var), shown for the cases
where it exceeds 90\%, and the binning of the lightcurves are reported
in Table~\ref{stats}.  They show that the full band lightcurve for the
2002 November observation was statistically consistent with a flat
count rate.  However, both a large excess in the standard deviation
and in the $\chi^2$ statistic show the 2001 September data to be
variable to very high probabilities (99.7 and $> 99.9\%$ likelihood,
repectively).  We investigated this variability further by splitting
each dataset into two arbitrary energy bands containing approximately
the same number of counts, a 0.3 -- 1.1 keV ``soft'' band and a 1.1 --
10 keV ``hard'' band.  The lightcurves for these bands are shown in
Figure~\ref{lcs}, again binned to the equivalent of an average of 25
counts per bin (c.f. Table~\ref{stats} for actual bin sizes).  Both
the tests on the binned lightcurves show no strong evidence for
variability in the soft band.  However, strong evidence for
variability is present in the hard band lightcurve in the 2001
September observation (at over 99.9\% likelihood in both tests), and
may also be present in the 2002 November observation ($\sim 96\%$
likelihood according to the $\chi^2$ test).

In order to provide a separate test of the variability free of any
gross photon binning effects we employed a Kolmogorov-Smirnov test, in
which we compared the cumulative photon arrival times to the expected
cumulative arrival times for a constant-flux source\footnote{Though
note that the data is strictly limited by the 3.24104 s temporal
resolution (which is the sum of the exposure time plus the frame
readout time) of the ACIS-S3 chip in the course of performing this
test.}.  This provided separate confirmation of the strong variability
in the 2001 September hard band data, and possible variability in the
2002 November hard data.  Interestingly, it also showed the 2002
November soft data to be variable at very high significance ($>
99\%$), which was not picked up by any other test.  An examination of
the photon arrival time distribution for the data shows this result is
due to a very slight increase in the average soft count rate over the
latter stages of the 2002 November observation, when compared to the
earlier stages, which the standard deviation and $\chi^2$ tests were
insensitive to.  The actual average count rate change was from $(5.25
\pm 0.17) \times 10^{-2}$ count s$^{-1}$ in the first 18.5 ks to
$(5.61 \pm 0.22) \times 10^{-2}$ count s$^{-1}$ in the latter 11.5 ks.
We note this is probably a real change, and is certainly not due to a
change in the background count rate for the source, which remains at
$\sim 10^{-4}$ count s$^{-1}$ in the soft band throughout the
observation.

Our statistical tests, and the amplitude of the variability shown by
Figure~\ref{lcs}, demonstrate that the majority of the hard X-ray
emission of NGC 6946 X-11 is varying at least on the timescale of the
binning size we use to maximise our statistics, $\sim 330 - 350$ s.
The hard X-ray emission is therefore originating in a region with a
light crossing time of less than six minutes ($\approx 10^8$ km).
This is consistent with a compact accretion-driven X-ray source such
as an X-ray binary, and not with a spatially-extended supernova
remnant.  Note, however, that we cannot currently exclude the majority
of the soft X-ray emission from originating in a more widespread
component.

\section{Discussion}

NGC 6946 X-11 is a point-like X-ray source, its X-ray spectrum is
well-fit by models used to describe ultraluminous BHXRB, and its hard
X-ray emission varies on timescales of less than six minutes.  The
simplest interpretation of this evidence is that the extraordinary
X-ray luminosity of NGC 6946 X-11 originates in the accretion of
matter in a BHXRB system, and not from the interaction of supernova
blast waves with circumstellar material, or a pulsar wind nebula, as
previously suggested.  Of course, we cannot exclude some spatial
extension in the X-ray source, which the lightcurves tell us would
have to be soft X-ray emission.  If we ignore for a moment the
unrealistically low column to the MEKAL plus powerlaw continuum model,
the MEKAL could still describe the hot gas phase of a supernova
remnant.  However, the spectral modelling implies that only $\la 10\%$
of the observed flux originates in the MEKAL component, implying that
the contribution to the observed luminosity from a hot plasma is
limited to $\sim 2.5 \times 10^{38} \ergsec$.

If we are dealing with an ultraluminous BHXRB, it appears quite
typical for the limited knowledge we have of this exotic class of
object.  A similar spectral model has been derived for the few other
ULX with reasonable quality X-ray spectral data (Miller et al. 2003;
Krauss et al. 2003; Kaaret et al. 2003), with the low inner-disc
temperatures for the MCDBB component interpreted as possible evidence
for the presence of intermediate-mass black holes.  We also note that
the dominant powerlaw continuum component has a value very similar to
that observed for Galactic BHXRBs in the high state ($\sim 2.5$;
c.f. Ebisawa, Titarchuk \& Chakrabati 1996).  One curious feature we
have identified is that the X-ray variability in this ULX is
predominantly above 1.1 keV, implying that it is the powerlaw
continuum spectral component that is varying, and not the thermal
emission from the accretion disc.  This exclusively hard X-ray
variability could quite conveniently explain the observed spectral
hardening with increased flux seen in many ULX (e.g. Fabbiano et
al. 2003).  However, the scenario they describe where the hardening is
due to a hotter inner accretion disc at higher accretion rates may not
be consistent with our observations, due to the short variability
timescales and the softness of the thermal disc component in the X-ray
spectrum.  The rapid variation of the hard X-rays may instead
originate in other physical mechanisms, for instance variations at the
base of a jet (c.f. Georganopoulos, Aharanion \& Kirk 2002), or
magnetic reconnection events in the accretion disc corona (c.f. Reeves
et al. 2002 for a discussion of the latter mechanism applied to the
quasar PDS 456).

Now that we have identified the X-ray emission from MF 16 with a
BHXRB, where does this leave the physical models for the extreme radio
and optical luminosity of the supernova remnant?  It may be possible
that the radio emission could originate from the BHXRB in a
relativistically-beamed jet, similar to that reported for an ULX in
NGC 5408 by Kaaret et al. (2003).  It is also quite possible that the
optical line emission may be a result of the presence of the BHXRB; we
note that nebulae have been reported to be present at or around the
positions of several ULX by Pakull \& Mirioni (2003a,b).  Some of
these nebulae are present as giant ($d \sim 100 - 400$ pc) bubbles
encircling the ULX position; they may be supernova remnants related to
the birth of the ULX, as could be the case around IC 342 X-1 (see also
Roberts et al. 2003), or may be inflated by the winds from young stars
or relativistic jets from the BHXRB.  Other nebulae, such as that
present near the Holmberg II ULX, shows signs of X-ray ionization.  We
note that two of the key diagnostics of X-ray ionization, the presence
of a He II $\lambda 4686$ emission line (c.f. Pakull \& Angebault
1986) and an abnormally high [O{\small III}]/H$\beta$ ratio
(Remillard, Rappaport \& Macri 1995) are present in the optical
spectrum of MF 16 presented by BFS01.  The similarity between MF16 and
other ULX nebulae is noted by Pakull \& Mirioni (2003b); we suggest
here that at least some of the extraordinary optical emission-line
luminosity of MF 16 may originate in X-ray excitation.  Furthermore,
we speculate that the nebula itself may be a young precursor of the
giant bubbles seen by Pakull \& Mirioni (2003a,b) around other ULX
systems.  Further studies of MF 16 in this context would be of great
interest.  Finally, we note that the \hst images of MF 16 suggest the
presence of a young stellar object at the centre of the nebula.  Young
stars or stellar clusters have been discovered coincident with at
least two other ULX (Goad et al. 2002; Liu et al. 2002), hence it is
possible that this object may be the optical counterpart of the BHXRB
itself.

\vspace{0.2cm}

{\noindent \bf ACKNOWLEDGMENTS}

The authors the referee, Eric Schlegel, for his useful comments that
helped to improve this paper.  We also thank Andy Ptak for assisting
us in the use of {\small XASSIST}.  TPR thanks Bob Warwick and Graham
Wynn for useful discussions, and PPARC for financial support.

\end{document}